\documentclass[twocolumn,superscriptaddress, prl]{revtex4-1}
\usepackage{amsfonts}
\usepackage{amssymb}
\usepackage{amsmath}
\usepackage{amsthm}
\usepackage{dsfont}
\usepackage{mathtools}
\usepackage{stackrel}
\usepackage{color}

\usepackage[pdftex]{graphicx,hyperref}

\usepackage{hyperref}

\newcommand{\ket}[1]{|#1\rangle}

\newcommand{\colvec}[2]{ \begin{pmatrix}#1\\#2\end{pmatrix} }
\newcommand{\rowvec}[2]{ \begin{pmatrix}#1 & #2\end{pmatrix} }

\begin{document}

\title{How universal is the entanglement spectrum?}

\author{Anushya Chandran}
\affiliation{Department of Physics, Princeton University, Princeton, NJ 08544}
\affiliation{Perimeter Institute for Theoretical Physics, 31 Caroline Street N, Waterloo, Ontario, Canada N2L 2Y5}

\author{Vedika Khemani}
\affiliation{Department of Physics, Princeton University, Princeton, NJ 08544}

\author{S. L. Sondhi}
\affiliation{Department of Physics, Princeton University, Princeton, NJ 08544}

\date{\today}

\begin{abstract}
It is now commonly believed that the ground state entanglement spectrum (ES) exhibits
universal features characteristic of a given phase. In this letter, we show that this belief
is false in general. Most significantly, we show that the entanglement Hamiltonian can undergo
quantum phase transitions in which its ground state and low energy spectrum exhibit singular changes, \emph{even} when the physical system remains in the same phase. For
broken symmetry problems, this implies that the ES and the Renyi entropies
can mislead entirely, while for quantum Hall systems the ES has much less universal content
than assumed to date.

\end{abstract}

\maketitle


\paragraph*{Introduction ---}

A set of ideas from quantum information has revitalized the
study of phase structure in condensed matter \cite{Amico:2008uq}. Amongst
these is the elucidation of the entanglement in wavefunctions,
especially ground states. Consider a bipartition of the system into two parts, $A$ and $B$. The reduced density matrix for $A$, $\rho_A = \mbox{Tr}_{B} |\psi\rangle \langle \psi|$  is obtained for the wavefunction $|\psi\rangle$ by tracing over the degrees of freedom in part $B$.
A frequently studied measure of entanglement is the entanglement entropy which is the von Neumann entropy of the reduced density matrix, $S = -\mbox{Tr}(\rho_A \ln \rho_A)$.
This quantity obeys an area law in gapped phases \cite{Eisert:2010ab} with a subleading universal correction indicating the presence of topological order \cite{Kitaev:2006aa, Levin:2006aa}.

In a striking development, Haldane and Li \cite{Li:2008pb} found more universal signatures in the largest eigenvalues of $\rho_A$. 
They defined the ``entanglement Hamiltonian" $H_E$ as $\rho_A = \exp(-H_E)$,
and its energy spectrum as the ``entanglement spectrum" (ES). 
They found that the low-energy ES of the Pfaffian quantum Hall state resembled the minimal edge excitation spectrum, and proposed the ES
 as a fingerprint of topological order. A large literature followed \cite{Bray-Ali:2009ve,Lauchli:2010jt,Yao:2010zl,Prodan:2010aa,Thomale:2010yf, Turner:2010eq,Fidkowski:2010dq, Pollmann:2010uq,Kargarian:2010fp,Hermanns:2011hc,Dubail:2011fu,Regnault:2011lq,Chandran:2011fv,Papic:2011cl,Hughes:2011oq, Qi:2012oq, Poilblanc:2012la,Swingle:2012ud}, and the idea has been applied to broken
symmetry \cite{Cirac:2011ys,Metlitski:2011uq,Alba:2012qd,Alba:2013vn, Kolley:2013mi,Poilblanc:2010fr,Humeniuk:2012rz,Pizorn:2013kq,James:2013ly} and near critical points \cite{Calabrese:2008aa,Sacramento:2011uo,De-Chiara:2012cr,Lepori:2013eu,Giampaolo:2013pd}. Broadly speaking, this body of work suggests that a) the low-energy
ES contains universal information about the
phase that goes beyond the entanglement entropy and b) this information reflects the actual excitation spectrum of the systems at issue.

In this letter, we offer a critique of these beliefs and show that the low-energy ES contains much less universal information than assumed. Define the canonical ensemble of $H_E$ as $\rho_{E} = e^{-H_E/T_E}$, where $T_E$ is the entanglement temperature. For any operator $O_A$ in $A$, $\langle O_A \rangle = \mbox{Tr}(\rho_A O_A) = \mbox{Tr} (e^{-H_E} O_A)$. \textit{Thus, all physical observables in the parent wavefunction are derived from the canonical ensemble of $H_E$ at entanglement temperature $T_E=1$.} On the other hand, the low-energy ES probes the limit
$T_E \rightarrow 0$. For generic Hamiltonians, these two
limits need not be in the same phase, and the exponentially fewer eigenstates near the ground
state contribute vanishingly to physical canonical averages at $T_E=1$. 

We show that, as a consequence, $H_E$ can exhibit quantum phase transitions (QPT) with accompanying
singular changes in the ES that are entirely spurious. 
These ES rearrangements take place away from actual phase boundaries and all \textit{physical} observables remain completely analytic. 
This implies that previously used diagnostics of phase structure based on the low-energy ES, such as the ``tower of states" in broken symmetry systems, quasi-degeneracies of the excitation spectrum and the entanglement gap as an order parameter can fail.
We first present simple, general arguments in the context of broken symmetries and then illustrate them by explicit computation for a 2D Ising model.
We then give an analogous treatment for the $\sigma_{xy}=1$ phase of the Chern insulator which will bring us full circle to the work of Li and Haldane.

A few results in the literature already suggest the need for caution in using the ES. The ES is a purely ground state property. More than one local Hamiltonian can have the same ground state and thus the same ES, and examples where the spectrum is gapped in one case and gapless in another are known \cite{Fernandez-Gonzalez:2012sy}. Further, by the area-law, $H_E$ lives in one dimension less than $H$ so the two spectra cannot easily match unless the low energy excitations are at the boundary. The failure mode of the ES discussed here is new and cuts across both considerations.  

\paragraph*{Systems with symmetry breaking ---}

The belief in the literature is that symmetry breaking in $H_E$ (and the related ``tower of states'' spectrum for continuous symmetries) reflects order in the underlying ground state. We show that this belief is mistaken.

Consider the Ising model in a transverse field $\delta$ (TFIM). In $d\ge 2$ dimensions, the TFIM is ferromagnetic for $\delta < \delta_c^P$ and paramagnetic for $\delta > \delta_c^P$. We use $P$ and $E$ to denote physical and entanglement related quantities. By the area law, the entanglement Hamiltonian associated with a cut in real space describes a $d-1$ dimensional boundary system.
The generic phase diagram of $H_E$ as a function of $T_E$ and $\delta$ (Fig.~1) must satisfy the following constraints: a) the $T_E = 1$ cut must coincide with the physical phase diagram, b) as $\delta \rightarrow 0$, $H_E$ projects onto the ideal ferromagnetic state and hence is ordered at all $T_E$, c) as $\delta \rightarrow \infty$, $H_E$ projects onto the ideal paramagnetic state and cannot support order at any $T_E$.
Together (a) and (b) imply the existence of the ordered region FM$_1$ wherein the boundary spins in $H_E$ exhibit ferromagnetic order; the bulk spins far from the cut will be trivially ordered at any $T_E$. Now (a) {\it also} implies that the boundary correlation length diverges as $\delta \rightarrow \delta_c^P$ at $T_E=1$.
The correlation length at $T_E=0$ is even longer at any $\delta$ near $\delta_c^P$. 
Thus, at small $T_E$, FM$_1$ should continue into a second ordered region FM$_2$ for $\delta > \delta_c^P$.
The phase FM$_2$ is spurious as, in this regime, the low-energy ES is ordered when the physical state is paramagnetic.
At $\delta_c^E>\delta_c^P$, there is a `pseudo' QPT from FM$_2$ into the paramagnet, accompanied by singular rearrangements in the ES.

\begin{figure}
\includegraphics[width=6cm]{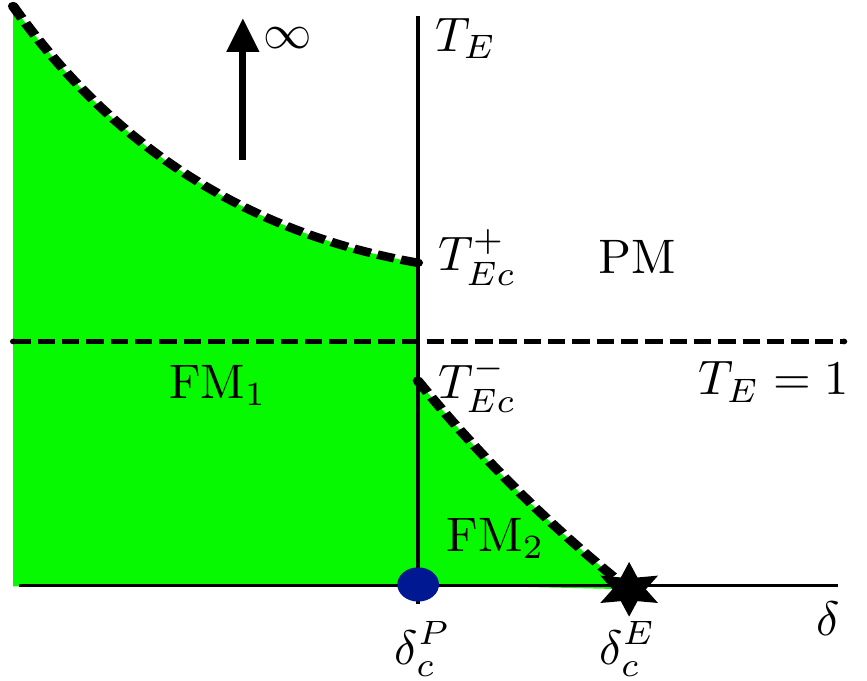}
\caption{Phase diagram of the entanglement Hamiltonian $H_E(\delta)$ for an Ising model. Blue dot at $\delta_c^P$: physical transition. Black star at $\delta_c^E$: pseudo QPT. FM$_2$: spurious ordered phase. Dashed line separating FM$_2$ and PM: pseudo transition in $d \ge 3$. If generic, $T_{Ec}^+=\infty$. }
\label{Fig: HEPhaseDiagram}
\end{figure}

A few comments about Fig.~\ref{Fig: HEPhaseDiagram} before we turn to a computation. 
First, {\it microscopic} couplings in $H_E$ exhibit singularities in the vicinity of the critical point $\delta_c^P$ \cite{Cirac:2011ys}. 
Thus, the two ordered phases on either side, FM$_1$ and FM$_2$, need not connect smoothly \footnote{In higher dimensions, the simplest hypothesis is that $T_{Ec}^- = 1$ on the critical line; the computations reported in Ref.~\onlinecite{Metlitski:2009bu} bear on what happens in $2<d <3$, but they currently deal only with subleading terms in the thermodynamics of $H_E$.}.
Second, for the generic TFIM, $T_{Ec}^+ = \infty$ as the boundary $H_E$ is always ordered when the bulk is. 
For the PEPS/Rokshar-Kivelson wavefunction we consider below, the evidence appears consistent with $T_{Ec}^+ < \infty$. 
Third, in $d=2$, by the Peierls-Mermin-Wagner theorem, FM$_2$ only exists at $T_E=0$.

We demonstrate the above scenario through an explicit calculation in a 2D Ising model. We work with the Rokhsar-Kivelson (RK) Ising wavefunction \cite{Rokhsar:1988kx},
\begin{align}
\ket{\Omega} = \sum_{\sigma} e^{-E_{cl}/2} \ket{\vec{\sigma}} \label{Eq:RK2d},
\end{align}
where $E_{cl}$ defines the classical anisotropic Ising model for spins $\sigma^z_{i,j} = \pm 1$ on sites $(i,j)$ of a 2D square lattice 
\begin{equation}
E_{cl} (\vec{\sigma}) = \sum_{i,j} -\beta_x (\sigma^z_{i,j} \sigma^z_{i,j+1}-1) -\beta_y (\sigma^z_{i,j} \sigma^z_{i+1,j}-1) \label{Eq:Eclassical}.
\end{equation}
The probability of a given configuration is $e^{-E_{cl}(\vec{\sigma})}$. Thus, the quantum RK wavefunction reproduces classical probabilities in the $z$-basis. The RK wavefunction is the ground state of a \textit{local} Ising-symmetric parent Hamiltonian $H_{RK}(\beta_x, \beta_y)$, which is quantum critical on the same critical line as the classical 2D Ising model \cite{Henley:2004ve,Castelnovo:2005zr,Ardonne:2004ly}: $\sinh(2 \beta_x^c) \sinh(2 \beta_y^c) = 1.$
This critical line defines the analog of $\delta_c^P$ in the preceding discussion.
\begin{figure}[htbp]
\begin{center}
\includegraphics[width=\the\columnwidth]{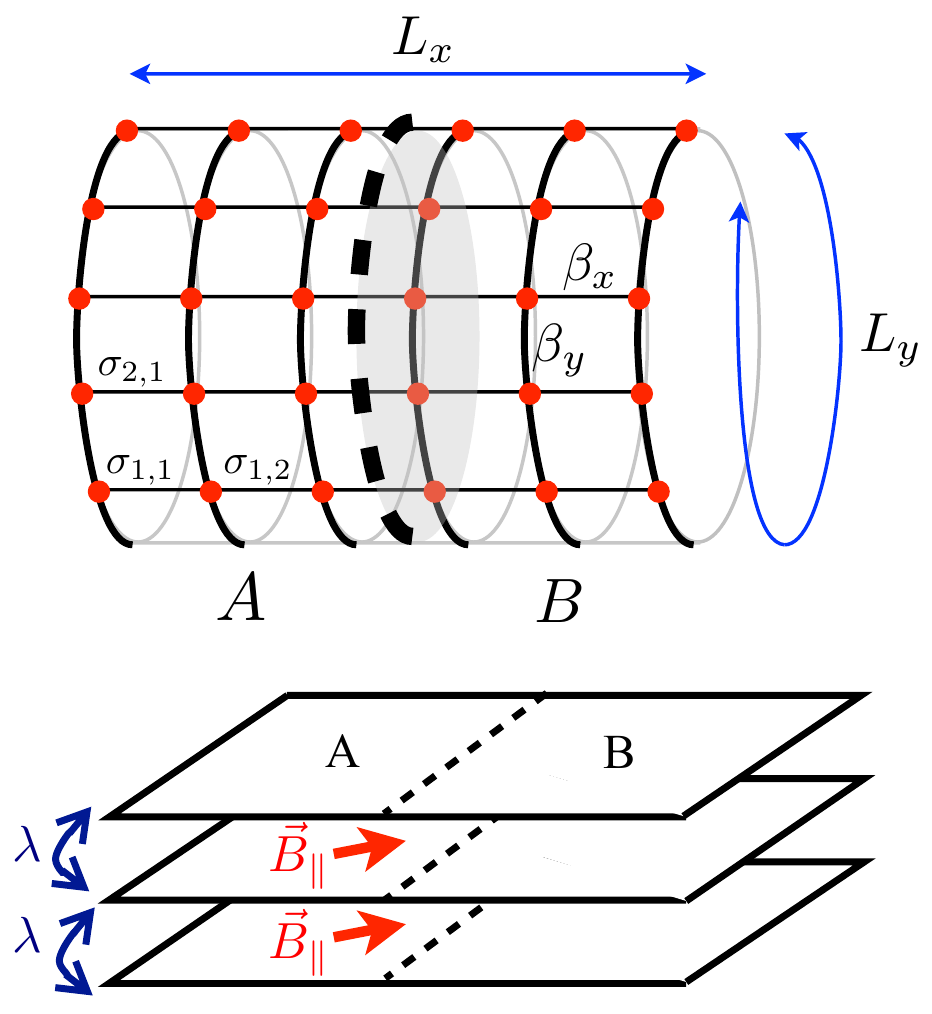}
\caption{Left: the RK Ising model on the square lattice. Right: the three layer Chern insulator model with magnetic flux between the layers. Region $B$ is traced out in the ground state to obtain the entanglement Hamiltonian in $A$.  }
\label{Fig:Models}
\end{center}
\end{figure}

To compute $H_E$, we place the system on an open cylinder of length $L_x$ and circumference $L_y$ and trace out half the cylinder (Fig.~\ref{Fig:Models}).  Consider the limit $\beta_x \gg 1$,  $\beta_y = 0$ in the PM phase. This corresponds to $L_y$ decoupled classical Ising chains parallel to the $x$ axis. As $|\Omega\rangle$ obeys a strict area law, the basis states of $H_E$ can be labelled by the spins at the boundary of the entanglement cut. For chain $i$, the two states are
\begin{equation}
\label{Eq:BulkEdgeLabel}
\ket{\sigma^z_{i,n}; A} \equiv \sum_{\substack {\sigma^z_{i,j}\\ \mathclap{j=0,\ldots n-1}} } \;e^{\sum_{j=0}^{n-1} \frac{\beta_x}{2} (\sigma^z_{i,j} \sigma^z_{i,j+1}-1)} \prod_{j=0}^{n}  \otimes\ket{\sigma^z_{i,j}} 
\end{equation}
where $n=L_x/2$ and $\sigma_{i,n}$ is the spin to the left of the cut. A small value of $\beta_y$ couples the chains. Using perturbation theory, we can systematically calculate $H_E$ in powers of $\beta_y$ (keeping $\beta_x \gg 1$). To first order,
\begin{align*}
 H_E
 = -2 e^{-\beta_x}\sum_{i=1}^{L_y}  \tilde{\sigma}_{i,n}^x - \frac{\beta_y e^{2\beta_x}}{2} \sum_{i=1}^{L_y} \tilde{\sigma}_{i,n}^z \tilde{\sigma}_{i+1,n}^z,
 \end{align*}
 where $\tilde{\sigma}^{x,z}$ act on the basis states in Eq.~\eqref{Eq:BulkEdgeLabel} in the usual way. The condition for our perturbative result to be valid is $\beta_y e^{2\beta_x} \ll 1 $. Although the states of $H_E$ are labelled by the boundary, they have weight in the 2d bulk.
 
We now present our central result. As $H_E$ is a 1D TFIM, it undergoes a pseudo QPT transition at a critical value of $\beta_y^E$
\begin{align}
\beta_y^E \sim 4 e^{-3\beta_x}.
\label{Eq:betacEnt}
\end{align}
This critical point lies within the regime of validity of our perturbation theory $\beta_y^E e^{2\beta_x} \ll 1 $. The physical transition in $H_{RK}$ is however at
\begin{align}
\beta_y^P \sim e^{-2\beta_x}.
\label{Eq:betacPhys}
\end{align}
Thus, $\beta_y^E \ll \beta_y^P$ with the inequality getting parametrically better for larger $\beta_x$; the zero temperature entanglement transition in $H_E$ precedes the physical one in a controlled limit. Essentially, the ground state of $H_E$ undergoes a surface ordering transition at $\beta_y^E$ before the bulk orders at $\beta_y^P$.

These pseudo transitions can be diagnosed by the oft-studied Renyi entropies. The Renyi entropy $S_n$ is proportional to $F_{1/n}-F_1$, where $F_T = -T \log(\textrm{Tr}\,e^{-H_E/T})$ is the entanglement free energy at temperature $T$. Therefore, whenever $H_E$ has a pseudo $T_E > 0$ phase transition, the $S_{1/T_E}$ Renyi exhibits unphysical singularities in the PM. In the $(2+1)$D TFIM, Singh et. al. \cite{Singh:2012ab} show that $S_2$ is analytic for $\delta\neq\delta_c^P$. This is consistent with the proposed phase diagram as only $S_\infty$ is singular in the PM phase in $d=2$. Previous Renyi studies of the RK Ising model have only focused on sub-leading terms \cite{Stephan2010ab}, while the relevant signature here is in the leading term. 

Recent studies in systems with continuous symmetry breaking \cite{Metlitski:2011uq,Alba:2013vn, Kolley:2013mi,Poilblanc:2010fr} have reported that the ES shows the characteristic ``tower of states" (ToS) structure of finite-size systems in the ordered phase. 
Our arguments show that the ES can be ordered even when the ground state is disordered in $d \ge 3$ (FM$_2$ in Fig.~\ref{Fig: HEPhaseDiagram}), with XY symmetric systems exhibiting Kosterlitz-Thouless order even in $d=2$. 
In this spurious region, the ES exhibits ToS structure, falsely diagnosing order.
If the symmetry is successively broken in stages (e.g. $O(4)\rightarrow O(3) \rightarrow O(2)$), then a spurious ToS may also appear in the putative ordered phase. 
However, we note that for non-RK wavefunctions, the spacing in the ToS in $H_E$ scales with system size $L$ as $e^{-L^{d-1}}$ in the spurious $FM_2$ phase, and as $e^{L^d}$ in $FM_1$.
In principle, this identifies the spurious ToS in this case \footnote{We thank M. Metlitiski and T. Grover for bringing this to our attention.}, though it requires more numerically intensive work. For RK wavefunctions, the scaling is the same in both ferromagnets.

\paragraph*{Chern Insulator/QH Hall fluid ---}

\begin{figure}[htbp]

\includegraphics[width=\columnwidth]{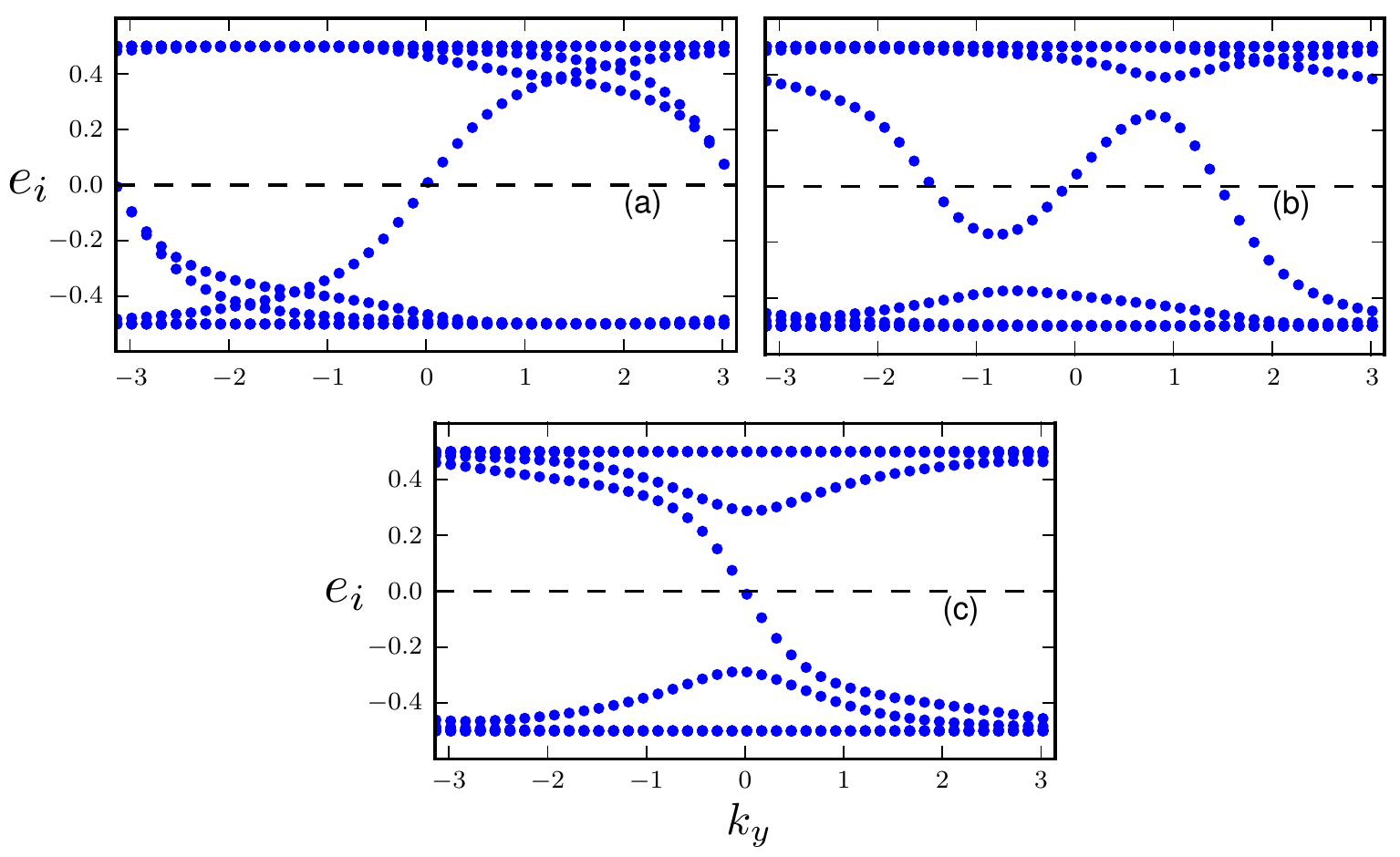}
\caption{Entanglement spectra (Eq.~\eqref{Eq:eiandxi}) of the ground states of the three layer model with $M_1 = -2.5$, $M_2=-1$, $M_3=-3.5$ and $L_x=L_y=200$. (a) $\lambda=0, \vec{A}_l=0$. (b) $\lambda=1/2$, $\vec{A}_1=-\vec{A}_3= (-\pi/2,-\pi/2)$, $\vec{A}_2 = (0,0)$. (c) $\vec{A}_1=-\vec{A}_3= (-\pi,-\pi)$, $\vec{A}_2 = (0,0)$.}
\label{Fig:Ex1ES}

\end{figure}

The quantum Hall (QH) fluids are incompressible in the bulk, but support chiral, gapless edge excitations.
The topological order in these fluids is manifest in the universal content of the edge theory.
In the ideal $\nu=1/3$ fractional QH Laughlin state, the edge is a single chiral boson with a universal compactification radius. 
The low-energy ES of such ideal states (and more realistic states) mimics the edge spectrum \cite{Li:2008pb}.

However, the edges of realistic states can exhibit edge reconstruction in which one or more non-chiral modes get added
to the spectrum \cite{Chamon:1994xu,Karlhede:1996if,Yang:2003uf}. This leads us to conjecture that the ES can {\it also} exhibit additional non-chiral modes. 
Further, there exist phase transitions in which the ES reconstructs while the system remains in the same topological phase. 
For related comments, see \cite{Swingle:2012ud}.
While the traditional discussion of edge reconstruction in the QHE requires interactions, we substantiate our conjecture below with an example involving a free fermion Chern insulator. 

The connection between the physical and the virtual edge is direct for free fermions \cite{Fidkowski:2010dq}. 
First, $H_E$ for the ground state of a free fermion Hamiltonian is quadratic \cite{Peschel:2003aa}\footnote{More generally, $H_E$ is quadratic for any Slater determinant many-body state.}. 
Thus, the ES vs the momentum along the cut, $k_y$ is a band spectrum. 
Second, the single particle entanglement energies, $\xi_i$, are monotonically related to $e_i$, the energies of a ``flat-band" Hamiltonian restricted to region $A$:
\begin{align}
e_i= (1/2) \tanh\left(\xi_i/2\right)
\label{Eq:eiandxi}
\end{align}
The ``flat-band" model on $A+B$ is in the same phase as the original Hamiltonian and has the same eigenstates, but with flattened bands \cite{Kitaev:2006bj}. Thus, the transitions at the physical edges of $A$ in the flat-band model appear in the ES of the ground state of $H$. We plot (Fig.~\ref{Fig:Ex1ES}) $e_i$ vs $k_y$ instead of the actual many-body ES for clarity.

Consider the $C=1$ Chern insulator on a periodic lattice. To allow for multiple edge modes, we take three independent bipartite layers in which the lower bands have $C=1$ in two layers and $C=-1$ in the third layer.
In the ground state, all three low energy bands are occupied and the system exhibits net $C=1$ with two right moving and one
left moving chiral edge modes, i.e. the edge content
exceeds the minimal edge content (central charge $c=1$) by one non-chiral
mode (whence $c=3$). This appears in the ES in Fig.~\ref{Fig:Ex1ES}(a) as well.
By deforming the Hamiltonian without closing the gap to the three high energy bands, we can modify the Chern numbers of the three lower bands to $C=1,0,0$.
The ES then exhibits one chiral mode with $c=1$ and there has to be a QPT in $H_E$ {\it en route}.

For simplicity, we give a different deformation in which we perturb the starting problem with a combination of interlayer hopping and uniform magnetic fields parallel to the layers (Fig.~\ref{Fig:Models}).
The Hamiltonian is: 
\begin{align}
H = \sum_{l=1}^3 h_l(\vec{A}_l, M_l) -\lambda \sum_{l=1}^2 &a^\dagger_{\vec{k}}(l+1) a_{\vec{k}}(l) + h.c.  \nonumber\\
&+ b^\dagger_{\vec{k}} (l+1)  b_{\vec{k}} (l) + h.c.
\label{Eq:Ham3Chern}
\end{align}
where $l$ is the layer index, $\lambda$ the hopping amplitude between identical sites on adjacent layers and
\begin{align*}
&h_l(\vec{A_l}, M_l) = \sum_{\vec{k}}\rowvec{a^\dagger_{\vec{k}}(l) }{ b^\dagger_{\vec{k}} (l) } (\vec{d}(\vec{k} + \vec{A_l}, M_l)\cdot \vec{\sigma}) \colvec{a_{\vec{k}}(l) }{ b_{\vec{k}} (l) } \\
&\vec{d}(\vec{k}, M_l) = \begin{pmatrix} \sin(k_x), & \sin(k_y), & 2+M-\cos(k_x)-\cos(k_y)\end{pmatrix}
\end{align*}
is the single layer Hamiltonian on layer $l$. Each layer is bipartite and $a^\dagger_{\vec{k}}(l), b^\dagger_{\vec{k}} (l) $ create Bloch waves on the two sublattices. $\vec{A_l}$ are constant
vector potentials corresponding to magnetic fluxes parallel to and
between the layers. The phase diagram of one layer at half filling is a function of $M_l$ only: for $M_l>0$ or $M_l<-4$, the ground state is a trivial insulator. When $-2<M_l<0$, the ground state is a Chern insulator with Chern number $C=-1$, while $C=1$ corresponds to $-4<M_l<-2$. At $M_l=0,-2,-4$, the system is gapless with Dirac fermion excitations.

At $\lambda=0, \vec{A}_l = \vec{0}$, we pick $M_l$ so that layers $1$ and $3$ are in the $C=1$ phase and layer $2$ is in the $C=-1$ phase. On turning on weak inter-layer coupling and magnetic fluxes, this point extends into a $C=1$ phase. The location of the Fermi points in the ES depends on a combination of $\lambda$ and the fluxes.
We can therefore arrange for the Fermi points of opposite chirality to be far apart in the starting configuration with the two left moving ones degenerate (Fig.~\ref{Fig:Ex1ES}(a)), then evolve into a configuration where they are all at distinct locations (Fig.~\ref{Fig:Ex1ES}(b)), and finally arrive at a configuration where an oppositely charged pair can meet and annihilate (Fig.~\ref{Fig:Ex1ES}(c)).

The transition in the ES is a genuine QPT, as the central charge changes. However, the net chiral central charge, $C_{c} = C_{left} - C_{right}$, is unchanged. More physically, there is a residual universality in  the edge (or ES) structures for a given bulk QH state as the conductance is the same at any $T$ (or $T_E)$. In principle, this can be extracted from the low-energy ES of sufficiently large systems (much larger than currently accessible in computational studies). All the methods to date to extract $C_c$ rely on the entire spectrum \cite{Zhang:2012ly,Cincio:2013gd,Zaletel:2013kl}.

We end with three comments. First, the quasi-degeneracies of particle-hole excitations as a function of momentum $\delta k_y$ relative to the
ground state of $H_E$---the commonly employed diagnostic for topological order--- change.
In Fig.~\ref{Fig:Ex1ES}(c), they take the values $\{1,1,2,3,5 \ldots\}$ for $\delta k_y=0,1,2 \dots$, while in the middle and left panels, they are modified to $\{1,2,5\ldots \}$ for $\delta k_y=0,1,2$.
This modified counting is not universal; by changing the speeds of the movers on the left edge, almost any sequence is possible.
Second, the arguments above should apply to fractional QH states, as they exhibit edge reconstruction. Recent work \cite{Cano:2013dw} has shown that for sufficiently complex abelian quantum Hall states, even the minimal edge structure is not unique and there can be phase transitions between distinct stable edge structures. The ES should exhibit analogous phase transitions. 
Third, while our general arguments apply to isotropic states, we work with anisotropic states for convenience. 

\paragraph*{Conclusions ---} This paper has two central messages about the low-energy entanglement spectrum. First, $H_E$ can exhibit spurious quantum phase transitions that have \textit{nothing} to do with any physical phase transitions. All physical observables are derived from $H_E$ at $T_E = 1$ and can remain analytic even as the low-energy ES exhibits singular changes.  Second, previously used diagnostics of phase structure based on the low-energy ES, such as the tower of states, quasi-degeneracies, and the entanglement gap, either fail completely or require much more careful analysis.  
Altogether our work indicates the need for caution in interpreting the results of ES computations.

\paragraph*{Acknowledgements ---}
We especially thank M. Metlitski and T. Grover for many discussions and for pointing out features of the phase diagram to us. We also thank J.M. Stephan, D. Haldane, A. Bernevig, N. Regnault and C. Laumann for comments on a draft of this article. This work was supported by NSF Grant Numbers DMR 10-06608, PHY-1005429
(AC, VK and SLS), the John Templeton Foundation (SLS), and the Perimeter Institute
for Theoretical Physics (AC). Research at Perimeter Institute is supported by the Government of
Canada through Industry Canada and by the Province of Ontario through the Ministry of Research and Innovation.

\bibliography{master}

\end{document}